\pdfoutput=1

\documentclass[11pt]{article}

\usepackage{acl}

\usepackage{times}
\usepackage{latexsym}

\usepackage[T1]{fontenc}

\usepackage[utf8]{inputenc}

\usepackage{microtype}

\usepackage{inconsolata}
 
\usepackage{times}
\usepackage{latexsym}
\usepackage{enumitem}
\usepackage{xcolor}
\usepackage{amsmath}
\usepackage{graphicx}
\usepackage{amsfonts}
\usepackage{comment}
\usepackage{booktabs}
\usepackage{url}
\usepackage{multirow}
\usepackage{subcaption} 

\newcommand{\jack}[1]{\textcolor{purple}{[jw: #1]}}

\newcommand{\model}{\texttt{Tempura}}

\title{Improve Temporal Awareness of LLMs for Sequential Recommendation}

\author{Zhendong Chu\textsuperscript{1}\thanks{\hspace{5pt}Work done during an internship at Adobe Research.}, $\,$ 
  Zichao Wang\textsuperscript{2}, $\,$ Ruiyi Zhang\textsuperscript{2}, $\,$ Yangfeng Ji\textsuperscript{1}, $\,$ Hongning Wang\textsuperscript{1}, $\,$ Tong Sun\textsuperscript{2} \\
  \textsuperscript{1}University of Virginia~~~~~~~~~~~~~~~~~~~~~~~~~~~~\textsuperscript{2}Adobe Research\\
\texttt{~~~~~~~\{zc9uy, yangfeng, hw5x\}@virginia.edu}, \texttt{\{jackwa,ruizhang,tsun\}@adobe.com}\\}

\begin{document}

\maketitle

\begin{abstract}

Large language models (LLMs) have  demonstrated impressive zero-shot abilities in solving  a wide range of general-purpose tasks. However, it is empirically found that LLMs fall short in recognizing and utilizing {\it temporal} information, rendering poor performance in tasks that require an understanding of sequential data, such as \emph{sequential recommendation}. In this paper, we aim to improve temporal awareness of LLMs by designing a principled prompting framework inspired by human cognitive processes. Specifically, we propose three prompting strategies to exploit temporal information within historical interactions for LLM-based sequential recommendation. Besides, we emulate \emph{divergent thinking} by aggregating LLM ranking results derived from these strategies. Evaluations on MovieLens-1M and Amazon Review datasets indicate that our proposed method significantly enhances the zero-shot capabilities of LLMs in sequential recommendation tasks. 


\end{abstract}

\section{Introduction}
Large language models (LLMs) such as ones with commercially available APIs including ChatGPT \cite{achiam2023gpt} and Claude\footnote{\url{https://www.anthropic.com/index/claude-2}}  have emerged as one of the primary, if not the de facto, choices in a wide range of applications thanks to their remarkable capabilities in dealing with natural language and generalizing to various domains without further fine-tuning. In deed, an emerging trend is to use natural language as a uniform interface and leverage the LLMs to complete a task.


\begin{figure}
    \centering
    \includegraphics[width=6cm]{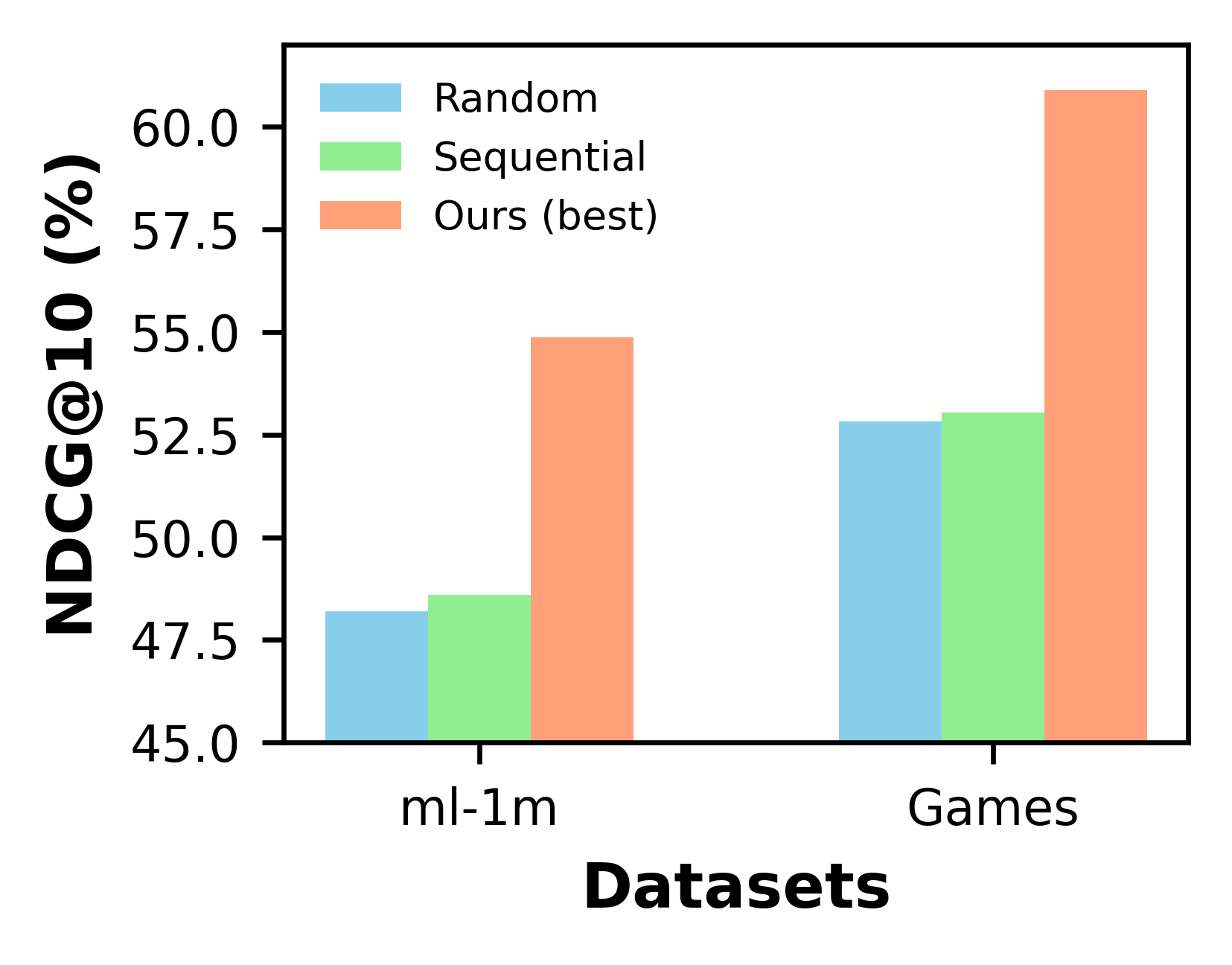}
    \caption{LLM-based sequential recommendation baselines show comparable performance even when historical interactions (Sequential) order is randomized (Random). {\model} significantly boosts performance by utilizing historical orders, \textit{i.e.}, temporal information.}
    \label{fig:motivation}
    \vspace{-3mm}
\end{figure}

Following this trend, recent research has been exploring the use of LLMs for processing sequential data, with applications such as sequential recommendation (SRS) \cite{hou2023large, bao2023tallrec}, which require LLMs to comprehend temporal patterns within user historical interactions.  In the case of sequential movie recommendation, historical interactions such as users' movie watching records can be represented as natural language (i.e., movie titles and other meta data) for the LLMs to process and recommend the next movie, instead of item identifiers which are typically used in traditional recommender systems \cite{kang2018self, sun2019bert4rec}. The extensive generalization ability and vast world knowledge \cite{wang2020language, singhal2023large} of LLMs endow them with the potential to serve as a single model for many recommendation domains without fine-tuning,  making it a general, capable, and easy-to-use alternative to traditional recommender systems that usually specialize in one selected domain and require extensive training or fine-tuning.



However, recent research shows that LLMs exhibit a limited sensitivity to temporal information in the input text, particularly in discerning changes in user interests \cite{hou2023large}. In Figure \ref{fig:motivation}, we compare the recommendation performance of LLM-based methods using randomized (denoted as \textbf{Random}) versus correctly ordered (denoted as \textbf{Sequential}) historical interactions on two widely-used SRS datasets. Both methods show similar performance, suggesting that LLMs are not effectively utilizing the temporal information present in the input text. This limitation stems from a lack of specialized mechanisms within LLMs to automatically recognize and utilize temporal information, which is crucial for understanding the context and progression within the data.

In this paper, we focus on improving LLMs' awareness and interpretation of temporal information, particularly within the SRS scenario. Temporal information is ubiquitous in real-world applications, such as recommender systems \cite{mcauley2022personalized}, intelligent document processing  \cite{fischer2001user} and financial market analysis \cite{tsay2005analysis}. By effectively capturing and integrating this temporal aspect, we have the opportunity to significantly enhance the understanding of user preferences via LLMs, thus providing users with better recommendations that suit their backgrounds, needs, and preferences. This improvement is also important for boosting the effectiveness of LLMs in downstream applications, where accurate user preference modeling is crucial \cite{mcauley2022personalized}. To this end, we design a principled prompting framework inspired by human cognitive process, which is training-free and domain agnostic. We name our approach as \textbf{\model} (phonetically similar to \textit{Temporal Prompt}). Our main contributions are:

\begin{itemize}[noitemsep, topsep=0.5pt,leftmargin=*]
\item  We propose a principled method to construct in-context examples \cite{min2022rethinking} for sequential recommendation, by analyzing how Transformer-based SRS models (e.g., \citet{kang2018self}) learn to utilize temporal information. 
\item Inspired by the results in neuroscience \cite{nobre2018anticipated, griffiths1998analysis}, we add explicit structure analysis in input sequences as additional prompts, particularly temporal cluster analysis, to enhance the temporal understanding capabilities of LLMs.
\item We emulate the process of \emph{divergent thinking} \cite{runco1991divergent} by aggregating ranking results derived from various prompting strategies.
\item We evaluate our method on MovieLens-1M and Amazon Review datasets, the results show that our proposed method significantly enhances the zero-shot capabilities of LLMs in sequential recommendation tasks. 
\end{itemize}

\section{Related Works}


\begin{figure*}
    \centering
    \includegraphics[width=14.2cm]{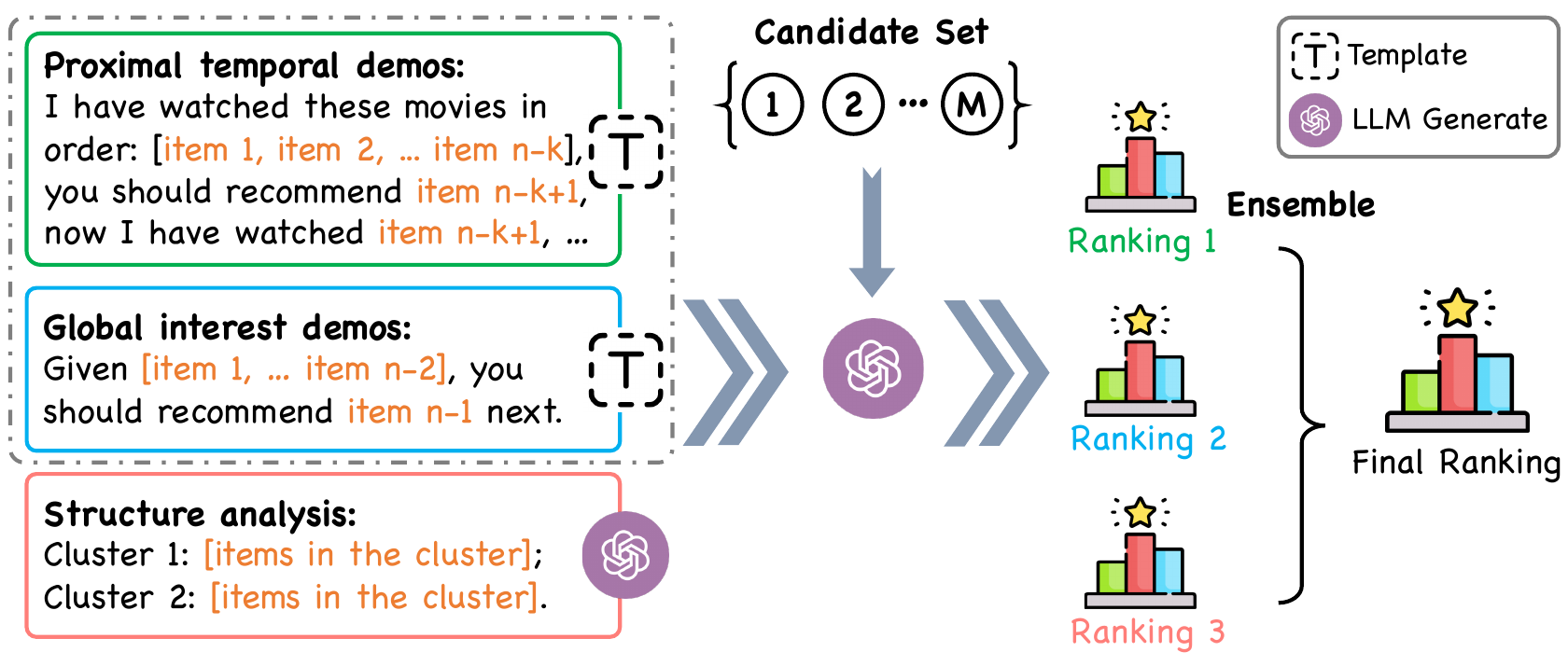}
    \vspace{-8pt}
    \caption{An illustrative overview of \model{}. We learn sequential recommendation via two kinds in-context demonstrations. Explicit  cluster structure analysis is conducted to improve the temporal understanding capabilities of LLMs. Each prompting strategy independently generates a respective ranking by LLMs (marked by different colors). Rankings from different prompting strategies are aggregated to form the final ranking.}
    \label{fig:overview}
\end{figure*}

\textbf{LLMs for recommendation.} Recently, the use of LLMs in recommendation systems has garnered significant research interest due to their capability to comprehend and encapsulate a user's preferences and past interactions through natural language \citep{fan2023recommender, he2023large}. Current LLM-based recommender systems are primarily designed for rating prediction \citep{kang2023llms, bao2023tallrec} and sequential recommendation tasks \citep{wang2023zero, hou2023large, xu2024prompting}. In both tasks, a user's previous interactions with items, along with other optional data like the user profile or item attributes, are concatenated to formulate a natural language prompt. This is then fed into an LLM with options for no fine-tuning \citep{wang2023zero}, full-model fine-tuning \citep{chen2023palr} or parameter-efficient fine-tuning \citep{bao2023tallrec}. \citet{liu2023chatgpt} designs a series of prompts to evaluate ChatGPT's performance over five recommendation tasks. \citet{wang2023recmind} develops a ChatGPT-based agent to improve recommendation ability by using tools such as SQL and Web search. Contrary to existing works that focus on the tentative evaluation of LLMs' ability in recommendation, we focus on improving the LLM's inefficacy of utilizing temporal information by designing temporal-aware prompting strategies.  



\noindent\textbf{Sequential recommendation.}  Sequential recommendation (SRS) \cite{hidasi2015session, kang2018self} aims to predict the next interacted items based on historical interaction sequences. Early works follow the Markov assumption \cite{rendle2010factorizing}, 
by designing various neural network models to capture user preference within interaction sequences, including Recurrent Neural Network \cite{hidasi2015session, li2017neural}, Convolutional Neural Network \cite{tang2018personalized}, Transformer \cite{kang2018self, sun2019bert4rec}, Graph Neural Network \cite{chang2021sequential, wu2019session}. However, most of these approaches are developed based on item IDs \cite{kang2018self} or attributes \cite{zhang2019feature} defined on specific domains, making it difficult to be generalized to other domains. Recently, \citet{hou2023learning}, \citet{hou2022towards} and \citet{li2023text} propose to learn unified item representations for SRS based on pretrained language models. They follow the paradigm that pretraining an unified text-based sequence encoder on source domains and then fine-tune the encoder on the target domain. However, all aforementioned methods need massive user interaction sequences on a specific domains and can not be easily transfer to unseen domains.  In contrast, we propose utilizing LLMs to establish a domain-agnostic learning process for sequential recommendation systems. Our approach is training-free and readily generalizable to unseen domains using only prompts. 
\section{Methodology}
In this section, we introduce \model{} in detail. As shown in Figure \ref{fig:overview}, \model{} consists of three major components: 1) a in-context learning module that learns sequential recommendation tasks from sequences of historical interactions; 2) a temporal structure analysis module that enhances the model's understanding by explicitly integrating cluster structures within the sequences; 3) a prompt ensemble module that aggregates recommendation results from various prompting strategies. We begin with the definition of notations to be used in our technical discussions. 

\subsection{Problem Definition}
Given a user's historical interactions $\mathcal{H} = \{i_j\}_{j=1}^n$, ordered chronologically up to timestamp $n$, the task of sequential recommendation involves ranking a set of \emph{candidate} items $\mathcal{C}=\{i_j\}_{j=1}^m$ for the subsequent timestamp $n+1$. Items of higher interest are expected to be ranked at more prominent positions. In practice, candidate items are typically selected from the entire item set $\mathcal{I}$, where $m \ll |\mathcal{I}|$, through candidate generation models \cite{covington2016deep}. Further, we follow the approach of \citet{hou2022towards} by associating each item $i$ with a descriptive text $t_i$, which could be the item's name and its attributes or properties. 

Different from training-based SRS models, we leverage general-purpose LLMs (e.g., ChatGPT) to solve the recommendation task in an instruction-following paradigm \cite{wei2021finetuned}. Specifically, for each user, we construct a history prompt from the user's historical interactions $\mathcal{H}$, and a candidate item prompt from the candidate item set $\mathcal{C}$. 
The aforementioned prompts are concatenated along with an instruction that explicitly describes the recommendation task, forming the final prompt for LLMs. LLMs are anticipated to generate rankings of $\mathcal{C}$, reflecting user preferences, in accordance with the format specified by the instruction. A post-hoc text parser is employed to convert the natural language rankings generated by LLMs into structured ranked lists, which is used to calculate the ranking metrics \cite{hou2023large}. 

\subsection{Sequential Recommendation via In-Context Learning}
\label{sec:prompts}
Given the vast scale of LLMs, fine-tuning domain-specific models becomes impractical. Thus, we propose to learn sequential recommendation via in-context learning, offering a training-free approach that can be easily adapted across various domains by leveraging the world knowledge and comprehension capabilities of LLMs \cite{hou2023large, harte2023leveraging}. To this end, we first analyze the learning process of training-based SRS models, and then mapping it onto the principles of constructing effective in-context demonstrations.


The key distinction between SRS and other recommender systems lies in the SRS model's requirement to not only identify a user's preferences based on historical user-item interactions but also to track the evolution of the user's interests over time. Training-based SRSs depend on learning from large-scale user-item interaction data via GRUs \cite{hidasi2015session} or Transformers \cite{sun2019bert4rec}. We utilize In-Context Learning (ICL) \cite{min2022rethinking} as a training-free alternative to learn a SRS model. 
We follow \citet{dai2022can} to analyze the learning process of training-based SRSs.  Given the historical interaction sequence of an user, a trained Transformer-based SRS, such as SASRec \cite{kang2018self}, can be represented as,
\begin{equation}
    \mathcal{F}_{\text{SASRec}}(\boldsymbol{x}_n) = (W_0 + \Delta W)\boldsymbol{x}_n.
\end{equation}
where $W_0$ is the initialized parameter matrix, $\Delta W$ is the update matrix and $\boldsymbol{x}_n$ is the representation of a candidate item. The output of $\mathcal{F}_{\text{SASRec}}$ is the score of the examined candidate item. In the back-propagation algorithm, $\Delta W$ is computed by accumulating the outer products of historic item representations $\boldsymbol{x}_i'^{T}$ and the error signals $\boldsymbol{e}_i$ of their corresponding outputs:
\begin{equation}
    \Delta W = \sum_{i=1}^{n-1}\boldsymbol{e}_i \otimes \boldsymbol{x}_i',
\end{equation}
where error signals $\boldsymbol{e}_i$ is the prediction error on the historic item $\boldsymbol{x}_i'$. Thus, the trained SASRec can also be rewritten into, 
\begin{align*}
    \mathcal{F}_{\text{SASRec}}(\boldsymbol{x}_n) &= (W_0 + \Delta W)\boldsymbol{x}_n \\
    &= W_0\boldsymbol{x}_n + \sum_{i=1}^{n-1}(\boldsymbol{e}_i \otimes \boldsymbol{x}_i') \boldsymbol{x}_n \\
    & = W_0\boldsymbol{x}_n + \text{LinAtt}(E, X', \boldsymbol{x}_n), \\
\end{align*}
where LinAtt($V, K, \textbf{q}$) denotes the linear attention operation, in which we regard error signals $E$ as values and interacted items $X'$ as keys, and the current input $\boldsymbol{x}_n$ as the query. The learning process of the SASRec model can be expained as the model predicting the next item in a sequence based on preceding items and updating itself based on the prediction error. The trained SASRec model is designed to update user preferences as the sequence expands, effectively tracking the evolution of the user's interests.

Let $\textbf{q} = W_Q\boldsymbol{x}_n$ represent the attention query vector for the input candidate item $\boldsymbol{x}_n$. An ICL-based SRS can be represented as, 
\begin{align*}
    \mathcal{F_\text{ICL}}(\textbf{q}) = (W_{\text{ZSL}} + \Delta W_{\text{ICL}})\textbf{q}
\end{align*}
where $W_{\text{ZSL}} = W_VX(W_KX)^T$ is the initialized parameters to be updated and $W_{\text{ZSL}}\textbf{q}$ is the attention result in zero-shot learning (ZSL) setting, where no demonstration are given. $X$ denotes the input representations of query tokens before $\boldsymbol{x}_n$, such as the task description of sequential recommendation. Based on the results of \citet{dai2022can}, the second term can be rewritten into,
\begin{align*}
    \Delta W_{\text{ICL}}\textbf{q} = \text{LinAtt}(W_VX', W_KX', \textbf{q}),
\end{align*}
where $X'$ denotes the input representations of demonstrations. Here we observe a similar form between $\mathcal{F}_{\text{SASRec}}$ and $\mathcal{F}_{\text{ICL}}$, where $W_VX'$ can be explained as the error signal from historic items. This analogy illustrates that by utilizing historic items as in-context demonstrations, an LLM can learn to capture the temporal information within the sequence of historical interactions. \citet{hou2023large} discussed using the last item in the history as an in-context demonstration. Based on our analysis, this method is equivalent to training the SASRec model solely with the last historical interaction, a practice insufficient for capturing the dynamic nature of historical interactions. Thus, we are motivated to use several historical interactions as demonstrations to improve the temporal awareness of LLMs. 

\noindent{\textbf{Proximal temporal demonstrations (PCL).}} Based on the above principle, we design the following prompt to learn to capture temporal information via ICL, 

\begin{figure}[!h]
    \vspace{-2mm}
    \centering
    \includegraphics[width=7.6cm]{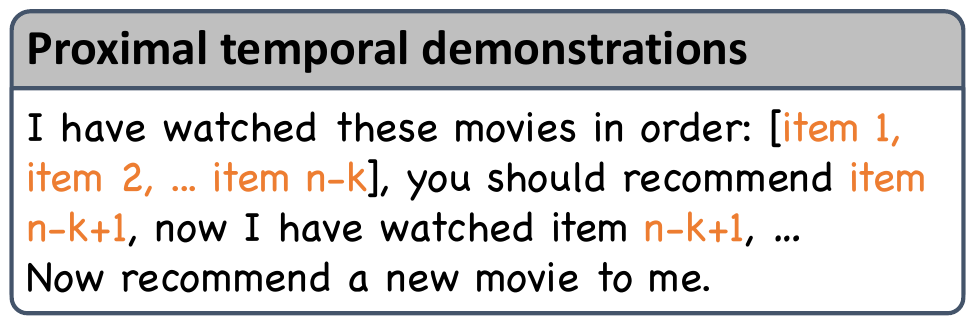}
    \label{fig:enter-label}
    \vspace{-3mm}
\end{figure}

Placeholders, highlighted in orange, structure the input for our model. The first placeholder captures the initial $n-k$ historic items, serving as the \emph{context} for inferring user preferences. The subsequent placeholder is designated for the $n-k+1$ item, illustrating the next item to be recommended based on the current context. Following this, we inform the LLM that the $n-k+1$ item has been interacted with, indicating that the $n-k+2$ item is the next recommendation target. This setup is repeated to create $k$-shot demonstrations. We utilize the most recent $k$ items as demonstrations to capture the proximal interest of the user. We denote this prompting strategy as PCL.

\noindent{\textbf{Global interest demonstrations.}} In previous studies \cite{kang2018self, hou2023large}, the number of historic items was constrained by the limited input length of models. Thus the whole interaction history is typically truncated and the most recent items are remained. Empirically, we also observed that extending the context window has limited impact on improving performance and may even detract from it. The reason could be: 1) the prolong context distract LLMs \cite{liu2023lost}; 2) too old history has little impact on the current user interest in the SRS scenario. However, simply omitting distant historic items risks overlooking users' long-term interests. Hence, we randomly sample a subset of historic items from the whole history sequence to retain user's global interest. Specifically, we use the same template as PCL, but the context is filled with randomly sampled historic items. Similarly, we incorporate the most recent items as in-context examples. We denote this prompting strategy as GCL.

\subsection{Temporal Structure Analysis} It has been recognized in the neuroscience area that the human brain is more sensitive to temporal structures \cite{nobre2018anticipated, griffiths1998analysis} - ``\emph{Embedded relationships among the attributes of events over different timescales carry predictions that guide proactive sensory and motor preparation in the brain}''. Only providing item sequences may make it difficult for LLM to identify and utilize temporal patterns inside the sequence. Thus, we are motivated to explicitly provide temporal structures to LLM. Specifically, we conduct cluster analysis on the item sequence according to two criteria: items that are (1) \emph{temporally proximate} and (2) \emph{share similar features} should be clustered. In practice, we also use LLMs to complete the cluster tasks and find it can provide reasonable cluster results. The results are used as additional input to the LLM for ranking. 

\begin{figure}[!h]
    \vspace{-3mm}
    \centering
    \includegraphics[width=7.6cm]{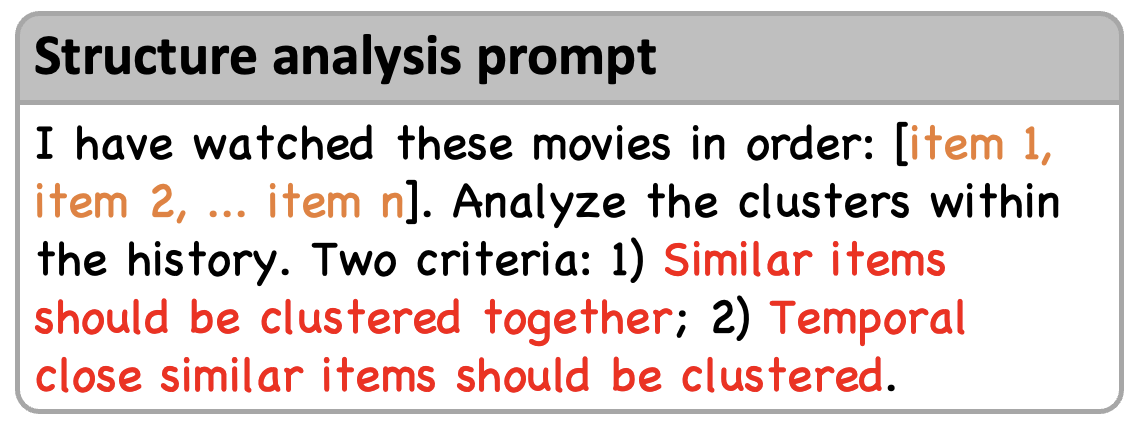}
    \label{fig:enter-label}
    \vspace{-3mm}
\end{figure}


\subsection{Prompt Ensemble} The most straightforward way to combine various prompting strategies is to concatenate them and use the resulted long prompt. However, this approach risks exceeding the context length limitations of LLMs. Moreover, it has been observed that LLMs may lose important information within overly lengthy prompts \cite{liu2023lost}. To effectively utilize different prompting strategies, we propose ensembling the respective ranking outcomes derived from each. In this approach, we create several LLM sessions and obtain ranking lists with different prompts. Following \citet{hou2023large}, we explicitly define the output format for the ranking results produced by LLMs, and subsequently extract the ranking list using a post-hoc text parser. These ranking lists are aggregated to obtain the final ranking, as the process shown in Figure \ref{fig:overview}. Existing research also highlights the benefits of collaboration among multiple LLMs \cite{wu2023autogen}. Specifically, we assign scores to each rank in the ranking list. For instance, in a ranking list of 20 items, the item in the 1st place receives 20 points, the 2nd place item gets 19 points, and so on, decreasing by one point per rank. Finally, we sum the scores for each item across all rankings.

\begin{table*}[!htp]
\centering
\caption{Performance comparison on ML-1M and Amazon Review datasets. We highlight the best performance in \textbf{bold}. N$@K$ denotes NDCG@$K$. }
\vspace{-8pt}
\label{tab:main}
\scalebox{1.0}{
\begin{tabular}{@{}ccccccccccccc@{}}
\toprule
\multirow{2}{*}{Method} & \multicolumn{3}{c}{ML-1M}     & \multicolumn{3}{c}{Games}  & \multicolumn{3}{c}{Kindle}    \\ \cmidrule(l){2-10} 
                        & N@1   & N@5   & N@10   & N@1   & N@5   & N@10 & N@1   & N@5   & N@10 \\ \midrule
BM25                    & 4.00  & 13.14 & 20.53   & 16.50  & 30.09 & 37.19 & 6.50  & 18.07 & 24.96  \\
UniSRec                 & 9.00  & 20.08 & 26.72  & 19.50 & 34.86 &  40.82 &  5.00  &  16.21 & 25.03  \\
VQ-Rec                  & 9.50  & 19.52 & 27.11  & 5.50  & 16.76 & 25.27 & 4.30 & 14.22 & 23.58 &  \\ \midrule
Sequential              & 21.43	& 42.57	& 48.59	 & 24.12 & 47.26 & 53.03 & 10.20 & 27.96 & 33.72  \\
RF                      & 26.56	& 45.99	& 51.27	 & 25.63 & 50.02 & 53.72 &  11.11 & 28.77 & 35.71  \\
ICL                     & 26.40 & 47.51 & 53.32  & 26.00 &	49.68 &	53.63 &  13.07 & 30.82 & 36.41 \\ \midrule
Cluster                & 27.00  &  45.82 & 52.04 & 26.15 & 47.41 & 52.39  & 13.20 &   25.77 & 34.07  \\
PCL                     & 29.16 & 48.44 & 54.21 & 29.00 &  51.56  & 55.11 &  11.55 & 29.45 & 36.46 \\
GCL & 30.50 & 48.53 & 53.26 & 32.00 & 51.61 & 56.63 & 10.00 & 31.45 &	36.67 \\ 
PCL + Cluster                   &  30.50 &  48.35 & \textbf{54.88} &  35.50 & 53.89 & 58.74 & 12.00 & 30.15 & \textbf{38.23}  \\ 
\model{}                   &  \textbf{31.50}  &  \textbf{48.64} & 54.49 &  \textbf{39.00} & \textbf{56.51} & \textbf{60.95} & \textbf{14.00} & \textbf{32.17} & 37.59  \\ \bottomrule
\end{tabular}
}
\vspace{-4mm}
\end{table*}

\section{Experiments}
In this section, to fully demonstrate the effectiveness of \model{} in improving temporal awareness of LLMs, we conduct a set of extensive experiments to study the following research questions: (1) Can \model{} improve LLM's performance on sequential recommendation compared to other methods?  (2) Can \model{} enhance the sensitivity of LLMs to temporal information in the input data? (3) How do factors like history length, the number of in-context examples or the choice of backbone LLMs influence the effectiveness of \model{}?

\subsection{Setup}
\noindent\textbf{Datasets.} The experiments are conducted on three widely-used public sequential recommendation datasets: (1) the movie rating dataset MovieLens-1M \cite{harper2015movielens} ({\bf ML-1M}) where user rated movies are regarded as interactions, (2) one category from the Amazon Review dataset \cite{ni2019justifying} named {\bf Games} where reviews are regarded as interactions, and (3) another category from Amazon Review dataset named \textbf{Kindle}. We sort the interactions of each user by timestamp, with the oldest interactions first, to construct the corresponding interaction sequences. The movie or product titles are used as the descriptive text of an item.

\noindent\textbf{Evaluation configurations.} Following existing works \cite{kang2018self, sun2019bert4rec, hou2023large}, we apply the leave-one-out strategy for evaluation. For each interaction sequence, the last item is used as the ground-truth item. We adopt the widely used metric NDCG@N to evaluate the ranking performance over the given candidate set $\mathcal{C}$ where $N\leq |\mathcal{C}|$. In the remainder of this paper, unless otherwise specified, $|\mathcal{C}|$ is set to 20. The candidate set consists of one ground-truth item and 19 randomly sampled negative items.

 \noindent\textbf{Baselines.} We consider three prompt-based baselines discussed in \cite{hou2023large}: \textbf{Sequential prompting}: Arrange the historical interactions in chronological order. \textbf{Recency-focused prompting (RF)}: In addition to the sequential interaction records, a sentence is additionally added to emphasize the most recent interaction. \textbf{In-context learning (ICL)}: Similar to PCL, but only use the most recent historic item as the in-context example. We also consider three methods designed for domain generalization: \textbf{BM25} \cite{robertson2009probabilistic} ranks items according to the textual similarity between candidates and historic items. \textbf{UniSRec} \cite{hou2022towards} equips textual item representations with an MoE-enhanced adaptor for domain fusion and adaptation. \textbf{VQ-Rec} \cite{hou2023learning}  learns vector-quantized item representations, which can map item text into a vector of discrete indices (i.e., item codes) and use them to retrieve item representations from a code embedding table in recommendations.  Additionally, we report the results with each single prompting strategy, as well as the results from ensembling PCL and cluster analysis.

 Training-based methods such as \citet{kang2018self, sun2019bert4rec} are not considered as baselines because: (1) They are designed based on item IDs, which can not be generalized to new domains with new ID spaces. (2) Our research focuses on improving the temporal awareness of LLMs, as evidenced by improved performance in sequential recommendations. Thus, our goal is not necessarily to develop a state-of-the-art sequential recommendation method. 

 \begin{figure}
    \centering
    \includegraphics[width=6.8cm]{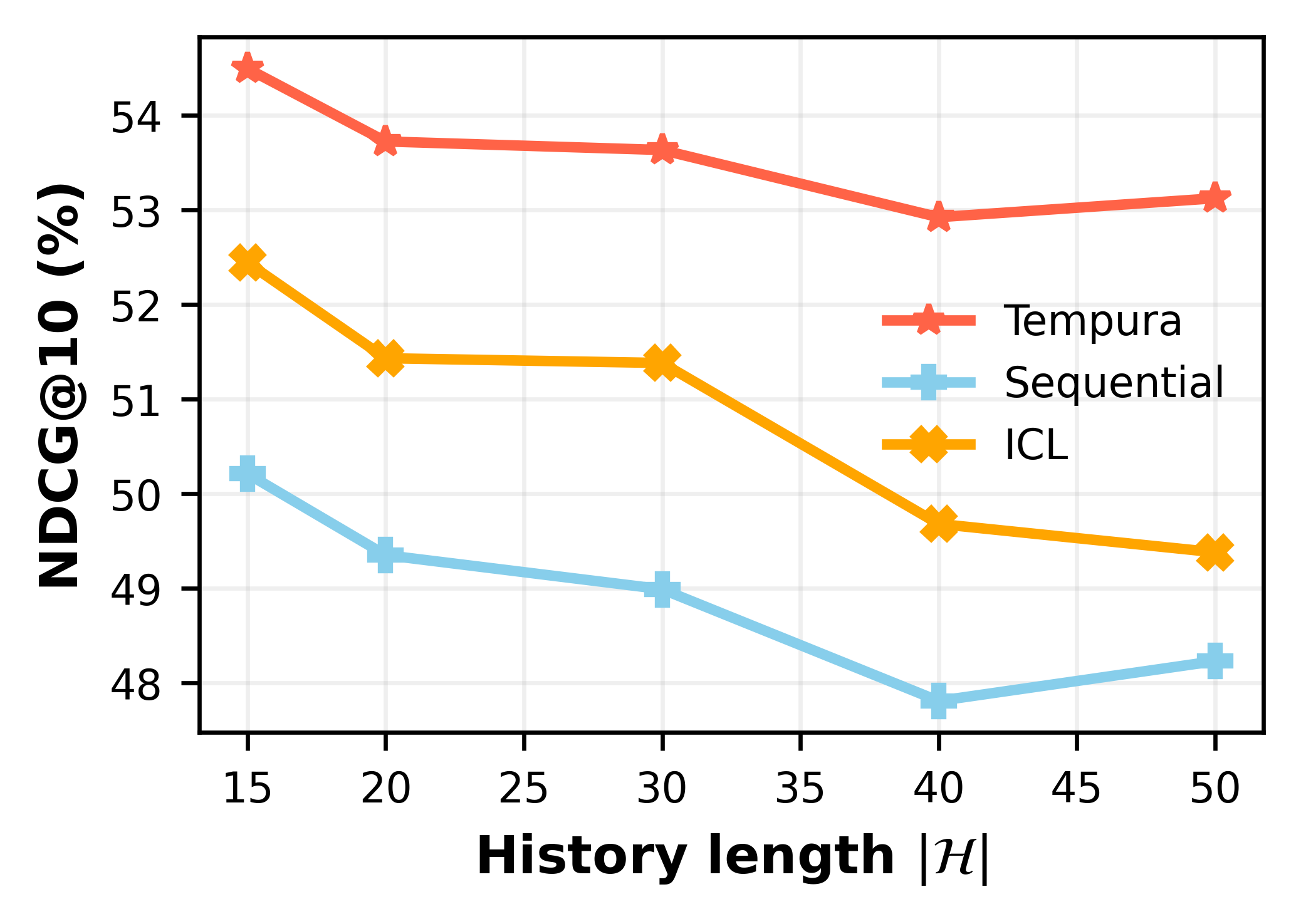}
    \caption{Performance vs. history length $|\mathcal{H}|$ (ML-1M).}
    \label{fig:history_length}
    \vspace{-3mm}
\end{figure}

\noindent\textbf{Implementation details.} Considering economic and efficiency factors, we follow \cite{hou2023large, xu2024prompting} to randomly sample 200 users along with their historical interactions for each dataset. Unless specified, we use the Azure OpenAI API \texttt{gpt-3.5-turbo}\footnote{\url{https://azure.microsoft.com/en-us/pricing/details/cognitive-services/openai-service/}}. We set history length $|\mathcal{H}|$ as 15 and use the most recent 5 interactions as demonstrations in PCL. We found the length of the history significantly affects performance; therefore, we also searched for the optimal $|\mathcal{H}|$ for baselines. Empirically, $|\mathcal{H}|=10$ yielded the best results for baselines in general. All the reported results are the average of three repeat runs to reduce the effect of randomness. 

\begin{table}[]
\caption{Performance of \model{} with randomized items, clusters and correctly ordered inputs. }
\label{exp:random}
\centering
\scalebox{0.9}{
\begin{tabular}{@{}cccc@{}}
\toprule
                           & Item-R & Cluster-R & Correct \\ \midrule
ML-1M & 51.78 &  52.47   &   54.49          \\ \midrule
Games & 51.83 & 54.18    &    60.95      \\ \midrule 
Kindle & 34.13  &  33.92    & 37.59      \\ \bottomrule
\end{tabular}
}
\vspace{-5mm}
\end{table}

\noindent\textbf{Main results.} We present the results on three datasets in Table \ref{tab:main}. We can observe our prompting strategies in the third group improves upon existing baselines across all metrics. It is interesting to observe that PCL outperforms ICL significantly, where more demonstrations are used in PCL but ICL only use the last interaction as demonstration. This observation align with our analysis that more demonstrations are needed to learn to utilize temporal information in historical interaction sequences. Although the Cluster strategy exhibits limited performance on its own, it can significantly enhance performance when combined with other strategies in an ensemble. Additionally, we provide a case study of cluster analysis results in Section \ref{sec:case_study}. By comparing individual prompting strategies with two ensemble-based methods, we find that ensembling consistently enhances performance by leveraging the strengths of different strategies. This suggests that different strategies emphasize various aspects, resulting in complementary results. 


\begin{figure}[!hp]
    \centering
    \includegraphics[width=6.8cm]{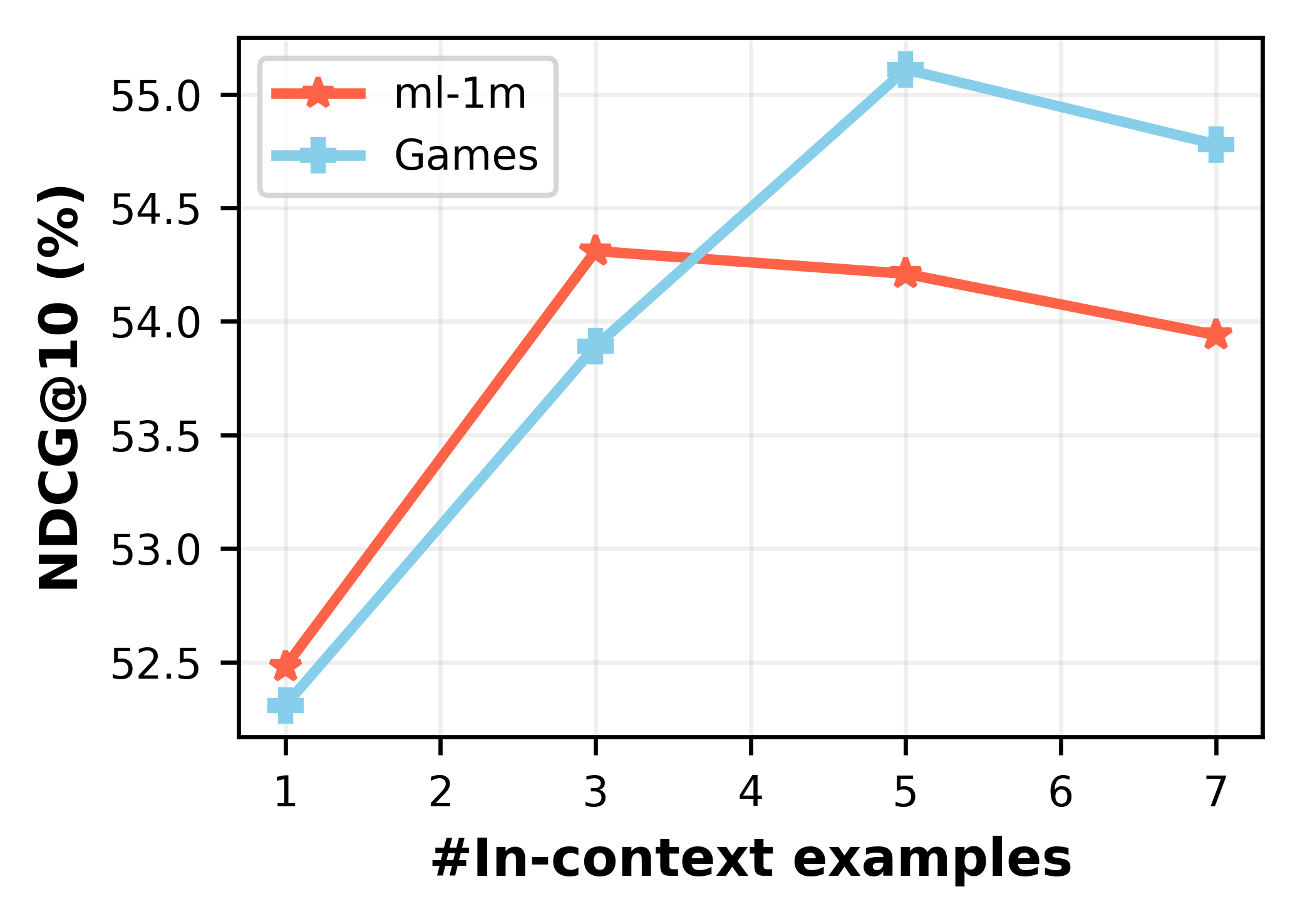}
    \vspace{-8pt}
    \caption{Impact of \#in-context examples in PCL. Several more examples can improve performance.}
    \label{fig:num_examples}
    \vspace{-3mm}
\end{figure}

\begin{table*}[!h]
\centering
\caption{Case study of structure analysis in the historical interaction sequence.}
\vspace{-8pt}
\label{tab:case_dis}
\begin{tabular}{@{}l@{}}
\toprule\toprule
\textcolor{red}{Cluster 1:} [Mad Max - PlayStation 4, Metal Gear Solid V: The Phantom Pain - PlayStation 4]. \\
\textcolor{blue}{Cluster summary:} Action games on PlayStation 4. \\ 
\textcolor{red}{Cluster 2:} [Star Wars: Battlefront - Standard Edition - PlayStation 4, Fallout 4 - PlayStation 4, \\ Just Cause 3 - PlayStation 4, Far Cry Primal - PlayStation 4 Standard Edition]. \\
\textcolor{blue}{Cluster summary:} Open-world action games on PlayStation 4. \\
\textcolor{red}{Cluster 3:} [Tom Clancy\'s The Division - PlayStation 4, Uncharted 4: A Thief\'s End - PlayStation 4, \\Homefront: The Revolution - PlayStation 4, Deus Ex: Mankind Divided - PlayStation 4]. \\
\textcolor{blue}{Cluster summary:} Action games with a focus on story and/or multiplayer on PlayStation 4. \\
\textcolor{red}{Cluster 4:} [Rise of the Tomb Raider: 20 Year Celebration - PlayStation 4, Dishonored 2 - PlayStation 4, \\ Resident Evil 7: Biohazard - PS4 Digital Code, Horizon Zero Dawn - PlayStation 4, Tom Clancy’s \\ Ghost Recon Wildlands - PlayStation 4]. \\
\textcolor{blue}{Cluster summary:} Single-player action shooting games with a focus on exploration and/or stealth on PS4. \\ \midrule
\textbf{Target item:} Prey - Pre-load - PS4 Digital Code \textcolor{orange}{First-person action-adventure shooting game}  \\
\bottomrule \bottomrule
\end{tabular}
\vspace{-3mm}
\end{table*}

\subsection{Sensitivity of Temporal Information}
In this paper, we aim to improve temporal awareness by designing temporal-aware prompting strategies. To evaluate whether these proposed strategies effectively capture and utilize temporal information within historical interaction sequences, we compare the performance with randomized and correctly ordered histories. We hypothesize that an approach adept at utilizing temporal information should demonstrate superior performance with correctly ordered history. Specifically, our manipulation occurs at two levels: the item level and the cluster analysis level. At the item level, we alter the order of individual items, while at the cluster analysis level, we rearrange the order of clusters derived from cluster analysis. We present the results in Table \ref{exp:random}. After randomizing the history, performance on all the datasets drop significantly. This phenomenon indicates that understanding and effectively utilizing temporal information within historical interaction sequences is crucial for capturing and predicting users' future interests.

\subsection{Ablation Study}

\noindent\textbf{Impact of history length. } It has been reported in \citet{hou2023large} that increasing the number of historical user behaviors does not improve the ranking performance, but even negatively impacts the ranking performance. To study the impact of history length on \model{}, we vary the history length $|\mathcal{H}|$ used for constructing the prompt from 15 to 50. We compare \model{} with the standard baseline Sequential and the best performing baseline ICL. Here history length $|\mathcal{H}|$ is the maximum allowed history length, the real history length could be shorter. We did not include the results on Games and Kindle since the user interaction history on these two datasets is short. 

The results are reported in Figure \ref{fig:history_length}. We observe that utilizing a longer history does not improve performance; in fact, it results in decreased performance on the ML-1M dataset. We hypothesize that the extensive history distracts LLMs, making it difficult for baselines to understand the evolution of user interests. By using temporal-aware prompts and the prompt ensemble strategy, \model{} demonstrates robust performance even with long historical interaction sequences.

\noindent\textbf{Impact of the number of in-context examples.} We utilize a user's historic items as in-context demonstrations to understand the temporal information in his / her behavior sequence. It is important to understand how many examples are needed. To this end, we study the performance with different number of examples in PCL. We keep the total length of the user's history as 15 and use the latest $k$ items as examples, setting $k$ to values in the set $[1, 3, 5, 7]$. We report the results on the ML-1M and Games datasets in Figure \ref{fig:num_examples}. We can observe more examples can boost the performance significantly than only one demonstration. As we analyzed in Section \ref{sec:prompts}, LLMs learn to utilize temporal information by learning to predict a series of historical items. However, it is not always the case that more is better. It is observed that a slight performance drop with more examples. We speculate that longer prompts may cause distraction for LLMs.

\noindent\textbf{Results on GPT-4.} More advanced LLMs, like GPT-4 \cite{achiam2023gpt}, demonstrate enhanced capabilities in knowledge, understanding, and reasoning. Therefore, we evaluate the sequential recommendation performance using GPT-4 to determine if \model{} can also augment GPT-4's capabilities. We present the results in Table \ref{tab:gpt4}. It has been observed that GPT-4 exhibits a robust capacity for sequential recommendation, even when employing the most standard prompting strategy, Sequential.  Notably, the improvement is most significant on the Kindle dataset, leading to the hypothesis that GPT-4 possesses extensive knowledge about Kindle books. The performance improvement with GPT-4 shows its strong ability in understanding and utilizing temporal information. By applying \model{}, the performance can be further improved when the backbone LLM is more powerful.   

\begin{table}[]
\centering
\caption{Performance with GPT-4 (NDCG@10). \model{} can further improve the performance when the backbone LLM is more powerful. }
\vspace{-1mm}
\label{tab:gpt4}
\begin{tabular}{@{}cccc@{}}
\toprule
Method     & ML-1M & Games & Kindle \\ \midrule
Sequential & 55.75 & 66.43 & 57.65  \\ \midrule
ICL   &    54.82    &  67.84   &   54.72                \\ \midrule
  \model{}         & 58.39 & 68.13 & 58.59  \\ \bottomrule
\end{tabular}
\vspace{-5mm}
\end{table}

\subsection{Case Study}
\label{sec:case_study}
We present an example result from the cluster analysis conducted on the Games dataset. We employ \texttt{gpt-3.5-turbo} to cluster historic items using the prompt discussed in Section \ref{sec:prompts}. The historic items was successfully clustered into 4 clusters, accompanied by a generated summary for each cluster. It can be easily observed that the user's most recent interest lies in  action shooting games. With this analysis, the target item can be easily identified since it is a first-person action-adventure shooting game, aligning with the user's latest interest.

\section{Conclusion}
In this paper, we focus on improving the temporal awareness of LLMs through the study of the sequential recommendation problem. Specifically, we introduce two kinds of prompting strategies: one to learn sequential recommendations via in-context learning and another to explicitly analyze the temporal structures in historical interaction sequences. An ensemble strategy is adopted to aggregate results from various prompting strategies. Our study demonstrates that by incorporating specific prompting strategies, LLMs can significantly improve in capturing and utilizing temporal information. This advancement not only strengthens the capabilities of LLMs in sequential recommendation tasks but also opens up new avenues for applying these models in time-sensitive domains.  

\section*{Limitations}
Firstly, although LLMs demonstrate notable capabilities in sequential recommendation, their performance still does not match that of training-based methods \cite{xu2024prompting, hou2022towards,kang2018self}. The reason could be that dataset-specific biases, which can be captured by trained models, are not inherently stored within LLMs. We leave combining prompt-based methods and training-based methods as an important future work. Secondly, considering the latency and costs in LLM inference, deploying LLM-based recommender systems could be cost-inefficient. However, LLMs provide a powerful and  explainable protocol to understand complex human behaviors in real-world recommender systems, making LLM-based systems a valuable complement to existing methodologies. 

\bibliography{anthology}

\end{document}